\documentclass[12pt, reqno]{amsart}
\usepackage{fullpage}
\usepackage[OT1]{fontenc}
\usepackage{amsthm,amsmath,amsfonts,amssymb,amsaddr}
\usepackage{mathtools}
\usepackage{hyperref}
\usepackage{xcolor}
\hypersetup{
    colorlinks=true,
    linkcolor=purple,    
    urlcolor=black,
    citecolor=blue
}

\usepackage{mathrsfs}
\usepackage{bm}
\usepackage{fancybox}
\usepackage{graphicx}
\usepackage{xcolor}
\usepackage{booktabs}
\usepackage{multirow}
\usepackage{subfigure}
\usepackage[round]{natbib}
\usepackage{algorithm}
\usepackage[noend]{algpseudocode}

\theoremstyle{definition}

\defcitealias{nycmta}{NYC MTA, 2020}

\newcommand{\iid}{\stackrel{\mathclap{\normalfont\tiny\mbox{iid}}}{\sim}}
\newcommand{\rise}{\mathrm{r}}
\newcommand{\dec}{\mathrm{d}}
\newcommand{\given}{\,|\,}

\title{Bayesian Hierarchical Modeling and Inference for Mechanistic Systems in Industrial Hygiene}

\author{Soumyakanti Pan$^{\star}$, Darpan Das$^{\dagger}$, Gurumurthy Ramachandran$^{\ddagger}$,\\Sudipto Banerjee$^{\star}$}

\address{$^{\star}$Department of Biostatistics, University of California Los Angeles, USA\\$^{\dagger}$Department of Environment and Geography, University of York, UK\\$^{\ddagger}$Department of Environmental Health Sciences and Engineering, Johns Hopkins\\Bloomberg School of Public Health and Whitmore School of Engineering, USA}

\begin{document}

\begin{abstract}
A series of experiments in stationary and moving passenger rail cars were conducted to measure removal rates of particles in the size ranges of SARS-CoV-2 viral aerosols, and the air changes per hour provided by existing and modified air handling systems. Such methods for exposure assessments are customarily based on \emph{mechanistic} models derived from physical laws of particle movement that are deterministic and do not account for measurement errors inherent in data collection. The resulting analysis compromises on reliably learning about mechanistic factors such as ventilation rates, aerosol generation rates and filtration efficiencies from field measurements. This manuscript develops a Bayesian state space modeling framework that synthesizes information from the mechanistic system as well as the field data. We derive a stochastic model from finite difference approximations of differential equations explaining particle concentrations. Our inferential framework trains the mechanistic system using the field measurements from the chamber experiments and delivers reliable estimates of the underlying physical process with fully model-based uncertainty quantification. Our application falls within the realm of Bayesian ``melding'' of mechanistic and statistical models and is of significant relevance to industrial hygienists and public health researchers working on assessment of exposure to viral aerosols in rail car fleets.

\end{abstract}
\maketitle

\section{Introduction}\label{sec:intro}

With the outbreak of the Covid-19 pandemic, public transit demand in the United States took a hit \citepalias{nycmta} as initial reports suggested it to be among the major vectors for transmission of the SARS-Cov-2 virus \citep{harris2020}. As it became clearer that the virus causing Covid-19 was transmitted via respiratory secretions which are aerosolized into tiny droplets \citep{Chia2020}, transit agencies took measures to minimize the exposure to the virus for passengers and employees. Following studies revealing inadequate social distancing rules in such settings \citep{bazbush21}, transit agencies have considered engineering interventions with the aim of reducing the risk of infection. While ventilation and filtration have always been integral to the air handling systems of train fleets, the Covid-19 public health crisis has brought increased attention on the effectiveness of engineering interventions. 

In partnership with a large-scale, interstate, mass-transit rail company in the United States, researchers have carried out a series of experiments inside a fleet of passenger rail cars sampled with a design accounting for various controls involving ventilation and filtration systems. The experiments focus on measuring concentration of aerosols at different locations inside the rail car compartment with an aerosol generator in the center. The aim is to ascertain important quantities related to ventilation and filtration system. As we do not observe the actual aerosol concentrations directly, but record partial noisy measurements, it is crucial from the industrial hygienist's perspective to understand the underlying physical process described by a system of deterministic differential equations.

Consolidating scientific inference by borrowing information from deterministic mechanistic systems and from field measurements designed to emulate the system continues to attract significant attention in diverse health science applications. Statistical approaches include Bayesian melding \citep[e.g.,][]{raftery95jasa, poole2000jasa, fuentes2005biocs, raftery2009biocs}, which synthesizes such information through a generic Bayesian hierarchical framework,
\begin{equation}\label{eq:generic}
    [\mbox{data}\given \mbox{process},\mbox{parameters}]\times [\mbox{process}\given \mbox{parameters}] \times [\mbox{parameters}]
\end{equation}
by modeling the field measurements (data), the mechanistic system (process) and all model parameters (mechanistic and statistical) jointly using probability distributions. Bayesian inference typically computes, or draws samples from, the posterior distribution of the process and parameters and carries out subsequent predictive inference by extending such inference to hitherto unmeasured observations. In its simplest form, Bayesian melding proceeds by regressing the data on the physical model. See, for example, \cite{zhang2009aoh} and \cite{raftery2010biocs} for two different applications. \cite{monteiro2014tch} demonstrate, however, that straightforward Bayesian nonlinear regression can be highly ineffective in predicting exposure concentrations in designed chamber experiments such as those encountered here. 

Using stochastic process emulators to model the output of the mechanistic system is widely used in calibrating computer models and similar approaches have been used in Bayesian melding \citep[see, e.g.,][]{monteiro2014tch, fuentes2005biocs}. In fact, such methods are often the only option when the mechanistic system is highly complex (e.g., climate models) and requires specialized computing environments for data analysis. In industrial hygiene, on the other hand, relatively simple differential equations comprise the mechanistic system which suggests building Bayesian dynamical systems for their analysis \citep{nada, wiklehooten, wiklecressie-R}. This enables mechanistic parameters to directly learn from the data obviating the need to carefully design runs, often multiple times, of the mechanistic system over a range of inputs. We work within such a paradigm here. 

The novelty of our application lies in the manner in which we address several data analytic challenges. First, the mechanistic models we consider incorporate multiple rise and decay of concentrations that are governed by the mechanistic parameters and experimental conditions. Assimilating this information requires a careful balance of statistical learning from the data as well as from the underlying deterministic mechanism. Second, we need to construct our inferential framework to handle streaming in as different cycles within the experiment. Industrial hygiene experiments typically involve a substantial amount of unreliable ``background data'' between cycles. We address this issue by allowing our framework to learn about the process in these background zones by assimilating mechanistic considerations with data driven inference. A specific contribution of this framework is aimed at public health researchers as we show the inferential benefits of performing an analysis by delving into the mechanistic equations over a black-box emulator-based inference based on multiple runs of the system.

The remainder of this manuscript evolves as follows. Section~\ref{sec:mechanistic_models} offers an account of different mechanistic models in industrial hygiene and provides scientific justification for our framework. Section~\ref{sec:experiment} describes the design and conduct of the field experiment. Section~\ref{sec:bayesian_modeling} develops the Bayesian hierarchical modeling framework while Sections~\ref{sec:simul}~and~\ref{sec:data_analysis} present analysis of simulated data and that of the field experiment, respectively. Section~\ref{sec:discussion} concludes the article with a discussion.  

\section{Mechanistic Models}\label{sec:mechanistic_models}
The ``one box model'' \citep{Reinke.Keil} is widely used in environmental engineering to assess occupational exposure when subject exposure occurs far from the source. The working assumptions of the model includes the ``well-mixed room'' assumption indicating a spatial uniformity of particle concentration inside the chamber at an instant. The assumption of the room being well mixed is due to either natural or induced air currents, which results in nearly equal concentration levels throughout the room.

The standard model assumes that a source is generating a pollutant at a constant rate $G$ in a room of volume $V$ and ventilation volumetric flow rate $Q$. The following differential equation describes the dynamics of particle concentration $C(.)$ inside the room, which is a function of time $t$. We will refer to this system as ``Model 101'' \citep{hewett1}
\begin{align}\label{eq:hew101}
    \text{Model 101:}\quad V \frac{dC}{dt} = G - CQ \;.
\end{align}
In practice, usually the generation is stopped after some time and the concentration eventually decays resulting in an experiment cycle. If the total time taken by an experiment cycle to end is $T$ with the generation stopped at time $T_0$, then the time dependent concentration during the exposure rise and decay of a cyclic process is given by the functions
\begin{equation}\label{eq:sol_hew101}
\begin{split}
    \text{Rise: } C_\rise (t; C_0, \phi) & = C_0 \ \mathrm{exp}\left(-\frac{Q}{V} t\right) + \frac{G}{Q} \left[ 1 - \mathrm{exp}\left(-\frac{Q}{V}t \right) \right],\quad t \leq T_0 \\
    \text{Decay: } C_\dec (t; C_0, \phi) & = C_\rise (T_0; C_0, \phi) \ \mathrm{exp} \left(-\frac{Q}{V} (t-T_0) \right),\quad t > T_0
\end{split} \;,
\end{equation}
where $\phi = (G, Q)$ denotes the unknown parameters of interest. Due to the inadequacy of Model 101 for exposure assessment in the presence of local engineering controls, \cite{hewett1} propose enriching the model with suitable parameters for local controls and develop a nested sequence of mechanistic models. The last, hence the richest, model in the sequence is described as ``one box, constant emissions, Local Exhaust Ventilation (LEV) with return, general ventilation with re-circulation" (acronym 1Box.CE.LevR.GvR). This model is applicable to a local exhaust setting in which the filtered air is returned to the workplace, but with an increase in the effective ventilation by the amount of recirculated air, accompanied by the efficiencies for contaminant collection, filtration etc. We refer to this model as ``Model 111'', which is described by the mass balance equation, 
\begin{equation}\label{eq:hew111}
    \text{Model 111:} \quad V \frac{dC}{dt} = (1-\epsilon_L \epsilon_{LF})G - C(Q + \epsilon_{LF} Q_L + \epsilon_{RF} Q_R)\;.
\end{equation}
The closed form solution of \eqref{eq:hew111} is a reparametrized version of the functions in \eqref{eq:sol_hew101}, $C_\rise (t; \phi^\prime)$ and $C_\dec (t; \phi^\prime)$ with $\phi^\prime = (G^\prime, Q^\prime)$ where, $G^\prime = (1-\epsilon_L \epsilon_{LF})G$ and $Q^\prime = Q + \epsilon_{LF} Q_L + \epsilon_{RF} Q_R$. Here, the parameters of interest are $\phi_1 = \{G, Q, Q_R, Q_L, \epsilon_L, \epsilon_{LF}, \epsilon_{RF}\}$. Table~\ref{tab:params} briefly explains the parameters involved in this mechanistic system.
\begin{table}[t]
    \centering
    \begin{tabular}{c p{7cm} l}
    \toprule
    Variable & Definition & Unit\\ \midrule
      $G$   & Generation rate & mg/min\\
      $V$   & Volume & $\mathrm{m}^3$\\
      $Q$   & Ventilation rate & $\mathrm{m}^3$/min\\
      $Q_L$ & local exhaust ventilation rate & $\mathrm{m}^3$/min\\
      $Q_R$ & room recirculation system ventilation rate & $\mathrm{m}^3$/min\\
      $\epsilon_L$ & fraction of the source emissions immediately captured
by the local exhaust & unitless (0,1)\\
      $\epsilon_{LF}$ & local exhaust return filtration efficiency & unitless (0,1)\\
      $\epsilon_{RF}$ & general ventilation recirculation filtration efficiency & unitless (0,1)\\
      \bottomrule
    \end{tabular}
    \caption{Hewett Model 111 parameters}
    \label{tab:params}
\end{table}

\section{Experiment}\label{sec:experiment}
Experimental investigations were carried out to measure the removal rates of particles in the size ranges of SARS-CoV-2 viral aerosols in three rail cars of the same fleet, representative of the rail company's most regularly used commuter passenger cars. Each rail car was 150.5 $\mathrm{m}^3$ (5,314 $\mathrm{ft}^3$) with a designed outdoor air intake flow rate of 34 $\mathrm{m}^3$/min and a designed total supply air flow rate of 102 $\mathrm{m}^3$/min. The air in the car is designed to be filtered 40.7 times per hour and replaced or changed with outdoor air 13.6 times per hour by the HVAC system. Outdoor air is brought into the rail cars' return air duct (return plenum) through dampers that regulate the airflow. Here, the outdoor air mixes with the recirculated air, passes through a MERV-8/13 filter,then moves through the heating and cooling elements before entering the supply air duct (supply plenum) to be distributed back into the car volume. An exhaust blower removes a portion of the cabin air to the outside depending on the position of a ventilation damper. 

\begin{figure}[t]
    \centering
    \includegraphics[width = 0.9\textwidth]{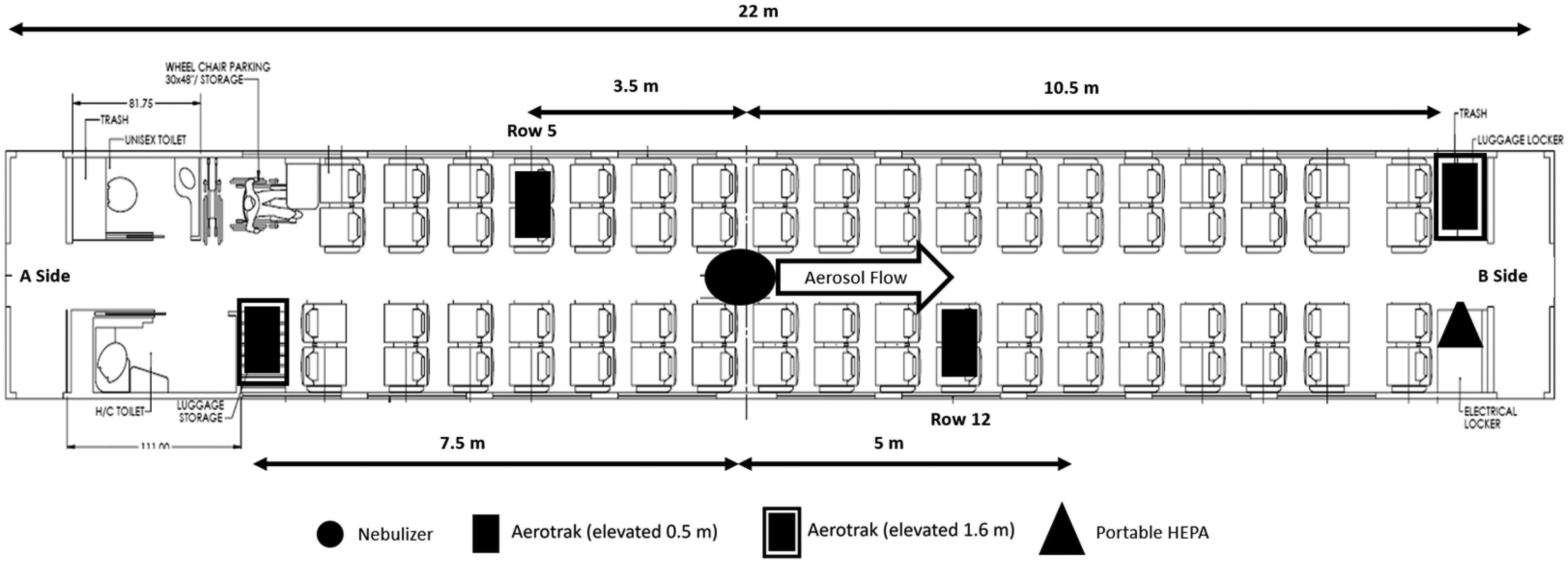}
    \caption{A schematic diagram of the experimental setup in a typical passenger car, drawn to scale with lengths in metres.}
    \label{fig:experiment}
\end{figure}

The rail cars can operate at speeds up to 201 km/h (125 mph). Each cabin had 36 seats on each side of a central aisle, spread over 18 rows, overhead compartments above each row, and two bathrooms on one end as shown in Figure~\ref{fig:experiment}.
Aerosols in the 0.3–5.0 mm size range were generated using a Collision nebulizer (MRE 3-jet with attached pressure gauge) with a 70:30 mixture of propylene glycol and vegetable glycerin. The nebulizer was placed in the center of the rail car between rows 10 and 11 (Figure~\ref{fig:experiment}), on a stand 1.0 m above the floor with the outlet 0.2 m above that. This height is equivalent to the distance from the floor to the middle part of the seat's headrest, making it a good  approximation for the  height of a person’s breathing zone and the origin of particle dispersion. 

Real-time aerosol concentrations were measured at four locations in the passenger cars using photo detector particle counters (AeroTrak Handheld Particle Counter- Model 9306; TSI; Shoreview, MN). The AeroTrak counts particles using a laser beam and a photodetector to detect light scattering and provides particle counts in six size ranges: 0.3–0.5 mm, 0.5–1.0 mm, 1.0–3.0 mm, 3.0–5.0 mm, 5.0–10.0 mm, and $>$ 10.0 mm. Each AeroTrak was calibrated daily, before beginning the experiments. Aerosol concentration measurements were logged at 1-min intervals for each experiment and downloaded to a computer as .csv files. Each experimental run consisted of 3 experiment cycles with each cycle carried out over a period of approximately 30 min with some background at the end, with the Collison nebulizer generating the aerosol for the first 15 min (aerosol concentration increase) and no aerosol generation for the second 15 min (aerosol concentration decrease). The intent was not to mimic human breathing or speaking but rather to observe the fate of aerosol particles of relevant sizes over time in the cabin. Complete details of the sampling instrumentation and experimental design are given in \cite{DasRam}.

\section{Bayesian modeling}\label{sec:bayesian_modeling}

The statistical model must account for the considerable amount of measurement errors and suitably quantify uncertainties in the field experiment. \cite{wiklehooten} offer a broad framework for statistical modeling exploiting knowledge of the underlying physical system available in the form of a dynamical system. We assume that a first-order Markov assumption is appropriate in this context and, hence, we introduce a process evolution model describing the latent true particle concentrations inside the chamber. 

The basic framework follows the model as given in \eqref{eq:obs.transn}. Due to a high degree of skewness in particle concentrations, it is reasonable to model the logarithmic concentration with Gaussian noise. Let $Y_t$ denote the measured concentration at time $t$ and let $C_t$ be the latent process representing the true concentration at time $t$. The observation equation allows the latent concentration to drive the inference while accommodating measurement errors. The transition equation models concentrations over time. These are formulated as
\begin{equation}\label{eq:obs.transn}
\begin{split}
    \text{Observation Equation: } \log Y_t &= \log C_t + \upsilon_t, \ \upsilon_t \sim P_\upsilon \\
    \text{Transition Equation: }\quad C_t & = f(C_{t-1}) + \omega_t, \ \omega_t \sim P_\omega\;,
\end{split}    
\end{equation}
where $\upsilon_t$ and $\omega_t$ are random processes modeling measurement errors and uncertainty in the concentration process through probability distributions $P_\upsilon$ and $P_\omega$, respectively, and $f(\cdot)$ is a specified function to introduce non-linearity in the transitions if needed. 

Replacing the instantaneous rate of change of concentration in \eqref{eq:hew101} by the average change in concentration in a time interval $(t, t+\Delta_t]$, yields an approximate relation between the concentration at the end and at the beginning of the interval. If $C(t+\Delta_t)$ is the underlying particle concentration at the next time point of measurements, with $\Delta_t$ being specified according to the time gap between successive measurements during the experiment and units of relevant parameters, we model the rise and decay as $\allowdisplaybreaks C_{t+1} \approx \left(1 - \frac{\Delta_t}{V} Q\right) C_t + \frac{\Delta_t}{V} G$ and $C_{t+1} \approx \left(1 - \frac{\Delta_t}{V} Q\right) C_t$, respectively. Therefore,
\begin{align}
        \text{Observation: } \log Y_t &= \log C_t + X_t^\top \beta + \upsilon_t, \ \upsilon_t \sim P_\upsilon \label{eq:ssm1} \\
        \text{Transition: } \quad C_t &= \left(1 - \frac{\Delta_t}{V} Q\right) C_{t-1} + \frac{\Delta_t}{V} G_t + \omega_t, \ \omega_t \sim P_\omega \label{eq:ssm2}
\end{align}
where, $X_t$ is a vector of explanatory variables at time $t$, $G_t = G {1}_\mathcal{G}(t)$, $1_{\mathcal{G}}(t)$ is the indicator function for $t\in \mathcal{G}$ and $\mathcal{G}$ is the collection of time points when the generation of particles was in place. The random process $\upsilon_t$ accounts for observation error and $\omega_t$ accounts for errors originating from the finite difference approximation of the differential equation and for possible biases in the deterministic model.


\subsection{Hierarchical Bayesian State-Space Model}\label{subsec:hierbayes}

\begin{figure}[t]
    \centering
    \includegraphics[width = 0.6\textwidth]{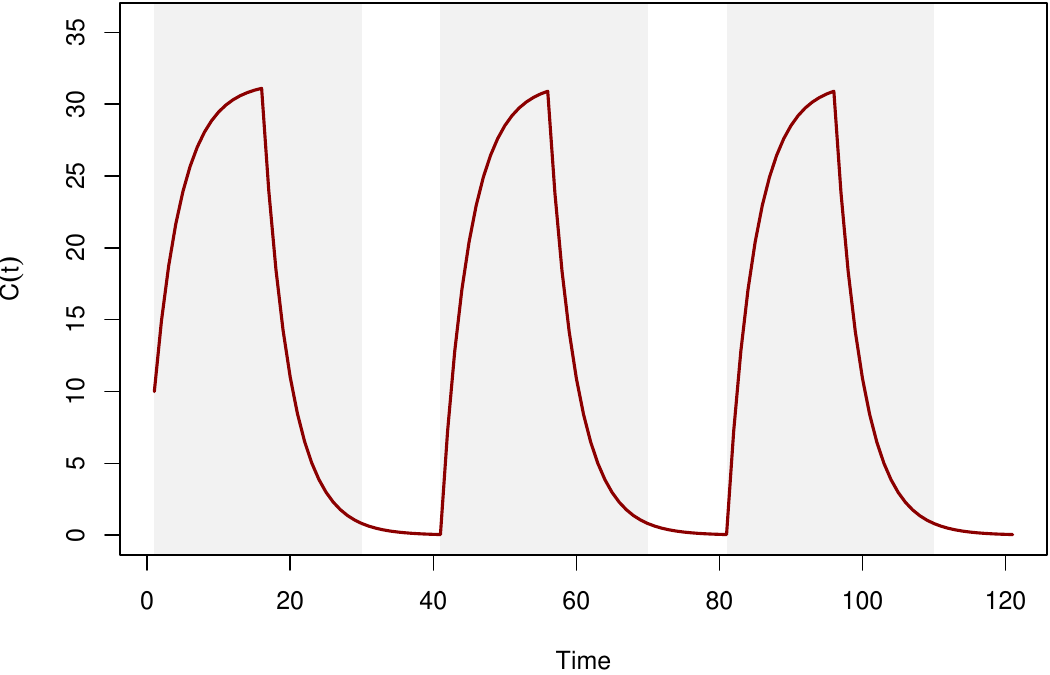}
    \caption{Plot of concentration curve $C(t)$ versus $t$ for a cyclic experiment with 3 cycles. The actual total length of the experiment is 120 minutes, with emission occurring during the first 15 minutes of each cycle and measurements are recorded only during the first 30 minutes of a cycle, marked with the shaded (in Grey) time intervals. The remaining data is treated as ``background''. The plot uses the following test values of the model parameters: $G = 1000$ mg/min, $V = 100$ $\mathrm{m}^3$, $Q = 20$ $\mathrm{m}^3$/min, $Q_L = 5$ $\mathrm{m}^3$/min, $Q_R = 5$ $\mathrm{m}^3$/min, $\epsilon_L = 0.6$, $\epsilon_{LF} = 0.3$, $\epsilon_{RF} = 0.9$, $C_0 = 10$ mg.}
    \label{fig:3cyc}
\end{figure}

We present a Bayesian state space model derived from \eqref{eq:ssm1} and \eqref{eq:ssm2} that address two challenges. First, the nature of the experiment generates consecutive cycles of data as described in Figure~\ref{fig:3cyc}. 
Second, each cycle is composed of both a rise and decay in concentrations as described in \eqref{eq:sol_hew101}, where the initial concentration of a cycle is derived from the estimated concentration in the second cycle. If $Z_t$ is the (possibly transformed) observed data and $g(\cdot)$ is a suitable transformation for the latent particle concentration at time $t$, $C_t$, then we construct the Bayesian dynamic model
\begin{align}\label{eq:hier}
\begin{split}
    Z_t\, &= g(C_t) + X_t^\top \beta + \upsilon_t, \ \upsilon_t \iid P_{\tau_1}, \\
    C_t\, &= A_t(\phi, \Delta_t)\, C_{t-1} + B_t(\phi, \Delta_t) + \omega_t, \ \omega_t \iid P_{\tau_2}, \\
    \{\phi, \beta, \tau \} \mid \psi & \sim p(\phi)\, p(\beta, \tau \mid \psi), \\
    \{\psi\} & \sim \pi(\psi)
\end{split}
\end{align}
where, $\tau = \{\tau_1, \tau_2\}$ are the parameters associated with the error distributions. The coefficients $A_t$ and $B_t$ in the process evolution are functions of $\phi$, the unknown parameters of the mechanistic model and the finite difference increments $\Delta_t$, which are known. In Model 101, $\phi = \{G, Q\}$ whereas, in Model 111, $\phi = \{G, Q, Q_R, Q_L, \epsilon_L, \epsilon_{L.F},\allowbreak \epsilon_{R.F}\}$. Usually prior information on the such parameters is scarce and, hence, uniform priors are considered. As we are modeling particle concentrations, it is reasonable to consider $g(\cdot)$ as the logarithm function and $P_{\tau_1}$ as the Gaussian distribution. 

Since the process evolution models particle concentration, we restrict $P_{\tau_2}$ to a distribution with non-negative support. A log-normal distribution for $P_{\tau_2}$ possibly dependent on time $t$ is a viable choice. \cite{nada} have used the Gamma distribution in mechanistic settings. Letting $Z_t = \log Y_t$, where $Y_t$ are the observed concentrations, we consider the following model incorporating mechanistic Model~111 in \eqref{eq:hew111}, 
\begin{align}\label{eq:uni.model}
\begin{split}
    Z_t \mid C_t, \beta, \sigma^2_\upsilon &\sim \mathcal{N}(\log C_t + X_t^\top \beta, \sigma^2_\upsilon) \\
    C_t \mid \phi , m_\omega, \sigma^2_\omega & \sim \mathrm{ShiftedLN} (A_t(\phi, \Delta_t)\, C_{t-1} + B_t(\phi, \Delta_t); m_{\omega}, \sigma^2_\omega) \\
    \{\phi, \beta, \sigma^2_\upsilon, m_{\omega}, \sigma^2_\omega \} &\sim p(\phi)\, p(\beta \given \sigma^2_\upsilon)\, p(\sigma^2_\upsilon)\, p(m_\omega)\, p(\sigma^2_\omega)\;,
\end{split}
\end{align}
where $A_t(\phi, \Delta_t) = 1 - (Q + \epsilon_{L.F} Q_L + \epsilon_{R.F} Q_R)\Delta_t/V$ and $B_t(\phi) = G_t\Delta_t/V$ with $G_t = (1-\epsilon_L \epsilon_{L.F})G 1_\mathcal{G}(t)$ as described in \eqref{eq:ssm2}. The random variable $X+\theta$ is said to be distributed as shifted log-normal $\mathrm{ShiftedLN}(\theta; \mu, \sigma^2)$ if $\log X$ is distributed normally with mean $\mu$ and variance $\sigma^2$ for some $\theta \in \mathbb{R}$. Setting $Q_L = Q_R = \epsilon_L = 0$ in \eqref{eq:uni.model} obtains a hierarchical model for Model~101 \eqref{eq:hew101}.


\subsection{Model for observed and latent states}\label{subsec:likelihood}

A salient feature of our analysis concerns the experiment being composed of $K$ cyclic experiments over the time period $\mathcal{T} = [0, T]$ with measurements taken over an ordered set of time points $0 < t_1 < t_2 < \cdots < t_N$, where $N$ is the total total number of observed time points. Recognizing that the background data (see Section~\ref{sec:mechanistic_models} and, more specifically, Figure~\ref{fig:3cyc}) collected between two cyclic experiments are often deemed unreliable, we estimate, with uncertainty quantification, the concentration state at the end of a cycle and use it as the assumed value at the start of of next cycle. We use the Bayesian hierarchical model in \eqref{eq:hier} to jointly model the observations and latent states over all the cycles.

Let $\mathcal{K} = \{t_1, t_2,\ldots, t_N\}$ be the set of time points at which the concentrations are measured over the duration of the experiment. We partition $\mathcal{K} = \sqcup_{i = 1}^K \mathcal{K}_i$ into $K$ distinct cycles, where $\mathcal{K}_i$ denotes all the time points generating measurements in cycle $i \in \{1,2,\ldots,K\}$ of the experiment and $\sqcup$ denotes disjoint unions. Let $t_i = \max \mathcal{K}_i$ be the last time point measuring concentrations for cycle $i$. Let $Y_\mathcal{K} = \{Y_{t_j} : t_j \in \mathcal{K}\}$ and $C_\mathcal{K} = \{C_{t_j} : t_j \in \mathcal{K}\}$ denote the sets of measurements and latent states of concentrations, respectively. The parameter space is given by $\Theta = \Theta_1 \sqcup \Theta_2$, where $\Theta_1$ and $\Theta_2$ are parameters present in the observation and latent equations, respectively. 

Building a hierarchical stochastic model for the observations and latent states conforming to \eqref{eq:uni.model} will need to account for the latent state at the end of a cycle as the value of the concentration state at the start of the next cycle is learned from the former. Let $\mathcal{S} = \{s_1,s_2,\ldots,s_K\}$, where $s_i$ denotes the starting time point of cycle $i$. We note that $s_i$ signifies the start of cycle $i$ and, therefore, is possibly distinct from the first time point in $\mathcal{K}_i$, which is the time point for the first measurement in cycle $i$. Therefore, $s_i \leq \min\mathcal{K}_i$. Assuming that the cycles are conditionally independent, given $\Theta$, the joint distribution of $Y_{\mathcal{K}}$ and $C_{\mathcal{K}}$ is 
\begin{align}\label{eq:lik}
    p(Y_\mathcal{K}, C_{\mathcal{K} \cup \mathcal{S}} \mid \Theta) = \prod_{i = 1}^K p(C_{s_i} \mid C_{u_{i-1}}, \Theta_2) \prod_{t_j \in \mathcal{K}_i} p(Y_{t_j} \mid C_{t_j}, \Theta_1) p(C_{t_j} \mid C_{t_{j-1}}, \Theta_2)\;,
\end{align}
where $u_i = \max\mathcal{K}_{i}$ denotes the end point of cycle $i$ and  $p(C_{s_1} \mid C_{u_{0}}, \Theta_2) = p(C_{s_1})$, which quantifies belief about the concentration state at the beginning of the first cycle, hence the starting condition of the experiment itself. 

The distributions $p(C_{t_j} \given C_{t_{j-1}}, \Theta_2)$ and $p(Y_{t_j} \given C_{t_j}, \Theta_1)$ are specified as shifted log-normal and log-normal, respectively, as in \eqref{eq:uni.model}. The parameters in \eqref{eq:sol_hew101} appear in \eqref{eq:lik} as $\Theta_1 = \{\beta, \sigma^2_v\}$ and $\Theta_2 = \{G, Q, Q_R, Q_L, \epsilon_L, \epsilon_{L.F}, \epsilon_{R.F}, m_\omega, \sigma^2_{\omega}\}$. We assign a log-normal distribution for $p(C_{s_i}\given C_{u_{i-1}},\Theta_2)$ such that $\log C_{s_i} \sim \mathcal{N}(\log \mu_{s_i},\sigma^2_{\omega})$, where $\log (\mu_{s_i}) = \log C_{u_{i-1}} - (Q/V)(s_i - u_{i-1})$ is derived from the mechanistic considerations embodied in \eqref{eq:sol_hew101}. Therefore, the latent concentration state at the beginning of a cycle learns from mechanistic considerations while also accounting for dispersion using the log-normal distribution. These specifications ensure a dynamic framework even as we marginalize over $C_\mathcal{S}$ leaving the distribution of the observed data dependent only on $\Theta$. Hence, for a fixed initial concentration $C_{s_1} = C_0$, \eqref{eq:lik} yields the joint distribution
\begin{align}\label{eq:marlik}
    p(Y_\mathcal{K}, C_\mathcal{K} \mid \Theta) &= \int \prod_{i = 1}^K p(C_{s_i} \mid C_{u_{i-1}}, \Theta_2) \prod_{t_j \in \mathcal{K}_i} p(Y_{t_j} \mid C_{t_j}, \Theta_1) p(C_{t_j} \mid C_{t_{j-1}}, \Theta_2) d C_\mathcal{S} \nonumber \\
    & = \prod_{t_j \in \mathcal{K}} p(Y_{t_j} \mid C_{t_j}, \Theta_1) p(C_{t_j} \mid C_{t_{j-1}}, \Theta_2)\;.
\end{align}
This reveals that the Markovian dependence within a cycle $\mathcal{K}_i$ in \eqref{eq:lik} is retained for any time point in $\mathcal{K}$. 


\subsection{Prior and Posterior}\label{subsec:prior_posterior}

We extend \eqref{eq:lik} to a joint distribution for $\{\Theta, C_{\mathcal{K} \cup \mathcal{S}}, Y_\mathcal{K}\}$ by specifying a prior distribution $p(\Theta)$. The posterior distribution is proportional to the joint distribution
\begin{align}\label{eq:posterior}
    p(\Theta, C_{\mathcal{K} \cup \mathcal{S}} \given Y_\mathcal{K}) \propto p(\Theta) \prod_{i = 1}^K p(C_{s_i} \mid C_{s_{i-1}}, \Theta_2) \prod_{t \in \mathcal{K}_i} p(Y_t \mid C_t, \Theta_1) p(C_t \mid C_{t-1}, \Theta_2)\;,
\end{align}
where the prior distribution corresponding to \eqref{eq:uni.model} is given by
\begin{equation}\label{eq:prior}
    \begin{split}
        p(\Theta) &=  \mathcal{N}\left(\beta \given \mu_\beta, \alpha \sigma^2_\upsilon\right) \times \mathcal{IG}\left(\sigma^2_\upsilon \given a_\upsilon, b_\upsilon\right) \times \mathcal{IG}\left(\sigma^2_\omega \given a_\omega, b_\omega\right) \times \mathcal{N}(m_\omega \given \mu_m, \kappa_m) \\
        & \quad \times \mathcal{U}\left(G \given a_G, b_G\right) \times \mathcal{U}\left(Q \given a_Q, b_Q \right) \times \mathcal{U}\left(Q_L \given a_{Q_L}, b_{Q_L} \right) \times \mathcal{U}\left(Q_R \given a_{Q_R}, b_{Q_R} \right) \\
        & \quad\quad \times \mathcal{U}\left(\epsilon_L \given a_{\epsilon_L}, b_{\epsilon_L} \right) \times \mathcal{U}\left(\epsilon_{L.F} \given a_{\epsilon_{L.F}}, b_{\epsilon_{L.F}} \right) \times \mathcal{U}\left(\epsilon_{R.F} \given a_{\epsilon_{R.F}}, b_{\epsilon_{R.F}} \right)\;,
    \end{split}
\end{equation}
where we denote $\mathcal{N}(X\given a, b)$, $\mathcal{IG}(X\given a, b)$ and $\mathcal{U}(X\given a,b)$ as Normal, inverse-Gamma and Uniform densities in $X$ with parameters $a$ and $b$, respectively \citep[][]{gelman2013}.


\subsection{Smoothing} \label{subsec:smooth}
A key inferential objective in dynamical systems is the smoothing of the latent process generating the data. In our current context, this amounts to model-based inference for the values of the latent concentrations at unobserved time points. Let ${\mathcal{Z}}$ be a finite collection of arbitrary time points where concentrations have not been measured. These points can be situated within the time duration of a cycle, a background time point for a cycle, or as a future time point of a cycle.

We use the posterior distribution $p(\Theta, C_\mathcal{K}\given Y_{\mathcal{K}})$ to evaluate the predictive distribution
\begin{align}\label{eq:smooth1}
    p(C_\mathcal{Z} \mid Y_\mathcal{K}) = \int p(C_\mathcal{Z} \mid Y_\mathcal{K}, C_\mathcal{K}, \Theta) \, p(\Theta, C_\mathcal{K} \mid Y_\mathcal{K}) \, d \Theta \, d C_\mathcal{K}\;.
\end{align}
Sampling from \eqref{eq:smooth1} is achieved as follows. For each value of $\{\Theta, C_{\mathcal{K}}\}$ sampled from $p(\Theta, C_\mathcal{K}\given \allowbreak Y_{\mathcal{K}})$, we draw one sample of $C_{\mathcal{Z}}$ from the conditional predictive distribution $p(C_\mathcal{Z} \mid Y_\mathcal{K}, C_\mathcal{K}, \Theta)$. Furthermore, we sample from the posterior predictive distribution of the measurements
\begin{align}\label{eq:smooth2}
    p(Y_\mathcal{Z} \mid Y_\mathcal{K}) &= \int p(Y_\mathcal{Z} \mid Y_\mathcal{K}, C_\mathcal{K}, \Theta) \, p(C_\mathcal{K}, \Theta \mid Y_\mathcal{K}) \, d \Theta \, d C_\mathcal{K} \nonumber\\
    &= \int p(Y_\mathcal{Z} \mid C_\mathcal{Z}, \Theta)\, p(C_\mathcal{Z} \mid Y_\mathcal{K}, C_\mathcal{K}, \Theta) \, p(C_\mathcal{K}, \Theta \mid Y_\mathcal{K}) \, d C_\mathcal{Z}\, d \Theta \, d C_\mathcal{K}
\end{align}
by drawing a $Y_{\mathcal{Z}}$ from $p(Y_\mathcal{Z} \given C_\mathcal{Z}, \Theta)$ for each sampled value of $C_{\mathcal{Z}}$ drawn from \eqref{eq:smooth1}. These samples provide full Bayesian inference for all points in $\mathcal{Z}$. If the points in $\mathcal{Z}$ lie within the domain of a cycle, the we obtain the smoothed values of the concentration state and measurements, while if the time points lie outside of the domain (in the future), we obtain forecasting estimates for the concentration state and predictions of measurements based upon values of the explanatory variables in $X_t$ at such points. 

\section{Simulation}\label{sec:simul}
We simulate three experiments. The first generates data from the mechanistic system described in \eqref{eq:hew101} using the parameter values $V = 100$ $\text{m}^3$, $G = 1000$ particles per minute and an average ventilation rate of $Q = 20$ $\text{m}^3/\text{min}$. We generated the data from the distribution of $Z_t$ in \eqref{eq:uni.model} setting $C_t$ to be the exact solution in \eqref{eq:sol_hew101} with $C_0 = 10$, $\beta=0$, $\sigma_\upsilon^2 = 0.01$ and $\Delta_t = 1$. We generate only one 20 minute cycle assuming that the particle generator is kept on for the first 15 minutes, which implies $T_0 = 15$ in \eqref{eq:sol_hew101}. The second experiment follows the same experimental specifications as the first but simulates 3 cycles three cycles. We assume that the particle generators are kept on for the first 15 minutes within each of the cycles, which implies that $T_0 = 15$ in the mechanistic system \eqref{eq:sol_hew101} for each of the three cycles. We generate the data over $90$ observed time points split into $\mathcal{K}_1 = \{1, \dots, 30\}$, $\mathcal{K}_2 = \{41, \dots, 70\}$ and $\mathcal{K}_3 = \{81, \dots, 110\}$. 

The third experiment changes the mechanistic alters the mechanistic system from the previous two. Here, we generate data for three cycles from the distribution of $Z_t$ in \eqref{eq:uni.model} using the mechanistic system in \eqref{eq:hew111} using $Q_L = Q_R = 5$ $\text{m}^3/\text{min}$, $\epsilon_L = \epsilon_{L.F} = 0.5$ and $\epsilon_{R.F} = 0.9$, while retaining the same parameter values for $V$, $G$, $Q$, $C_0$, $\beta$ and $\sigma_\upsilon^2$ as in the first and second experiments. The sets of indices at which data are observed is same as that of the second experiment. We analyze these data using \eqref{eq:uni.model}; see Section~\ref{subsec:simul_results}. 

\subsection{Priors for mechanistic parameters in simulation experiments}\label{subsec:priors}

Recall that our model parameters are classified into $\Theta_1 = \{\beta, \sigma^2_{\upsilon}\}$ representing parametric linear regression coefficients and a measurement error variance, and $\Theta_2 = \{\phi, m_\omega, \sigma^2_{\omega}\}$, where $\phi$ denotes the parameters in the mechanistic model under consideration. For the most general model in \eqref{eq:hew111}, we have $\phi = \{G, Q, Q_R, Q_L, \epsilon_L, \epsilon_{LF}, \epsilon_{RF}\}$ while \eqref{eq:hew101} has $\phi=\{G,Q\}$. We use the family of priors specified in \eqref{eq:prior} with $a_G$ = 200, $b_G$ = 1800, $a_Q$ = 3, $b_Q$ = 50, $a_{Q_L}$ = 2, $b_{Q_L}$ = 10, $a_{Q_R}$ = 2, $b_{Q_R}$ = 10, $a_{\epsilon_L}$ = $a_{\epsilon_{LF}}$ = 0.3, $b_{\epsilon_L}$ = $b_{\epsilon_{LF}}$ = 0.7, $a_{\epsilon_{RF}}$ = 0.6, $b_{\epsilon_{RF}}$ = 1, $a_{\upsilon}$ = 10, $b_{\upsilon}$ = 8.42, $a_{\omega}$ = 2, $b_{\omega}$ = 1.68, $\mu_m$ = 0 and $\kappa_m$ = 100.

Priors for the set of mechanistic parameters $\phi$, which are involved in the process evolution can be defined completely by the user or can be derived from the heuristic methods often followed by the experimenter to get rough estimates of the parameters \citep[see, e.g., the model calibration procedure in][]{hewett1}. The methods can include considering the log-transformed concentration only for the decay part of an experiment and regressing them on time. In case of \eqref{eq:hew101}, the regression coefficient of time yields estimates of the ventilation rate $Q$, which, in turn, will provides estimates for $G$ when the log-transformed concentration of the rise in \eqref{eq:sol_hew101} is regressed on time. For more complex models, such as \eqref{eq:hew111}, these heuristic methods fail to estimate all the parameters involved with ventilation. Other engineering interventions are necessary to overcome these problems, where they exploit the nested nature of the models. \cite{hewett1} remarks that calibration procedures are akin to back-of-the-envelope calculations for practicing occupational hygienists. However, these calculations can be used to build reasonable priors for parameters of $\phi$. 

\subsection{Computation} \label{subsec:computation}
All models discussed here are implemented in \texttt{R}~4.3.1 using \emph{rjags} \citep{rjags}. The posterior inference for each model is based on Markov chain Monte Carlo (MCMC) chains with 5000 iterations retained after discarding the initial 5000 samples as burn-in. These programs were executed on a single Apple M1 chip, with 3.20 GHz base clock speed and 8 GB of random-access memory running \texttt{macOS Ventura} (Version 13.4.1). We assessed convergence of MCMC chains by visually monitoring autocorrelations and checking the coverage of parameter estimates (posterior mean and 95\% credible interval) with the true values for the simulated data. Codes and data required to reproduce the results and findings in this article are openly available at \href{https://github.com/SPan-18/RailcarExposuremodelling}{Github} (active link for downloading). 

\subsection{Simulation results}\label{subsec:simul_results}
\begin{figure}[t]
\centering
\includegraphics[width=0.8\textwidth]{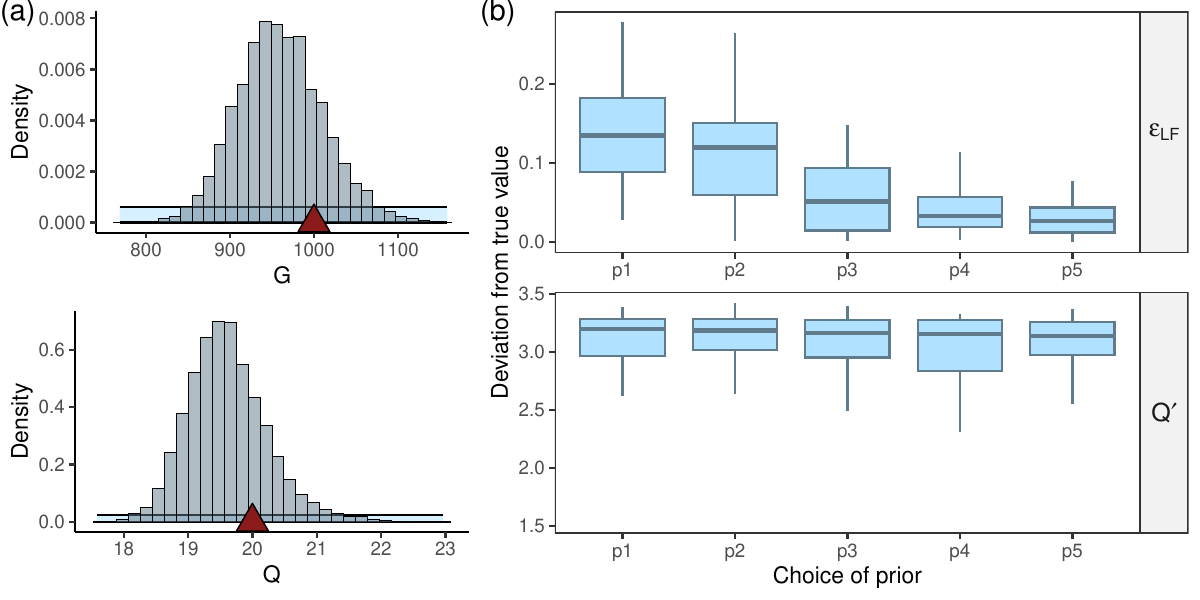}
        \caption{Posterior learning of mechanistic model parameters---(a) The histograms represent the posterior samples of $G$ and $Q$ while the light blue density denotes the prior of the respective parameters; (b) Deviation of the posterior median estimates from its true values for the parameter $\epsilon_{LF}$ and $Q' = Q+ \epsilon_{LF} Q_L + \epsilon_{RF} Q_R$ for different choices of priors. The priors p1 through p5 denotes different uniform priors on the mechanistic parameters, ordered according to increasing information in the prior. While more precise priors effectuate a decrease in the deviation of the posterior median estimate from its true values for the unidentified parameter $\epsilon_{LF}$, we see no such pattern for the posterior learning of $Q'$.}
     \label{fig:post.learn}
\end{figure}

\begin{table}[t]
\centering
\resizebox{0.95\textwidth}{!}{%
\begin{tabular}{@{}ccclcclcclcc@{}}
\toprule
\multicolumn{1}{l}{\multirow{3}{*}{Model}} &
  \multirow{3}{*}{\begin{tabular}[c]{@{}c@{}}Number\\ of \\ cycles\end{tabular}} &
  \multirow{2}{*}{\begin{tabular}[c]{@{}c@{}}Smoothing by\\hierarchical\\model\end{tabular}} &
   &
  \multicolumn{8}{c}{Smoothing by INLA} \\ \cmidrule(l){5-12} 
\multicolumn{1}{l}{} &
   &
   &
   &
  \multicolumn{2}{c}{B-splines} &
   &
  \multicolumn{2}{c}{Cubic splines} &
   &
  \multicolumn{2}{c}{\begin{tabular}[c]{@{}c@{}}Random walk\\models of order 2\end{tabular}} \\ \cmidrule(lr){3-3} \cmidrule(lr){5-6} \cmidrule(lr){8-9} \cmidrule(l){11-12} 
\multicolumn{1}{l}{} &
   &
  WAIC &
   &
  Knots &
  WAIC &
   &
  df &
  WAIC &
   &
  Type &
  WAIC \\ \midrule
\multirow{7}{*}{\begin{tabular}[c]{@{}c@{}}Model\\ 101\end{tabular}} &
  \multirow{3}{*}{1} &
  \multirow{3}{*}{-39.1} &
   &
  5 &
  -19.0 &
   &
  3 &
  -16.4 &
   &
  RW2 &
  -19.7 \\
 &
   &
   &
   &
  8 &
  -17.9 &
   &
  10 &
  -18.2 &
   &
  CRW2 &
  -19.4 \\
 &
   &
   &
   &
  20 &
  -12.6 &
   &
  20 &
  -26.2 &
   &
  \multicolumn{1}{l}{} &
  \multicolumn{1}{l}{} \\
 &
  \multicolumn{1}{l}{} &
  \multicolumn{1}{l}{} &
   &
  \multicolumn{1}{l}{} &
  \multicolumn{1}{l}{} &
   &
  \multicolumn{1}{l}{} &
  \multicolumn{1}{l}{} &
   &
  \multicolumn{1}{l}{} &
  \multicolumn{1}{l}{} \\
 &
  \multirow{3}{*}{3} &
  \multirow{3}{*}{-38.8} &
   &
  8 &
  -55.4 &
   &
  7 &
  -12.3 &
   &
  RW2 &
  -121.9 \\
 &
   &
   &
   &
  12 &
  -81.6 &
   &
  15 &
  -85.1 &
   &
  CRW2 &
  -119.6 \\
 &
   &
   &
   &
  24 &
  -10.8.5 &
   &
  20 &
  -96.2 &
   &
  \multicolumn{1}{l}{} &
  \multicolumn{1}{l}{} \\ \midrule
\multirow{7}{*}{\begin{tabular}[c]{@{}c@{}}Model\\ 111\end{tabular}} &
  \multirow{3}{*}{1} &
  \multirow{3}{*}{-17.2} &
   &
  5 &
  -18.1 &
   &
  3 &
  -16.8 &
   &
  RW2 &
  -19.9 \\
 &
   &
   &
   &
  8 &
  -17.8 &
   &
  10 &
  -17.3 &
   &
  CRW2 &
  -19.5 \\
 &
   &
   &
   &
  20 &
  -12.3 &
   &
  20 &
  -26.8 &
   &
  \multicolumn{1}{l}{} &
  \multicolumn{1}{l}{} \\
 &
  \multicolumn{1}{l}{} &
  \multicolumn{1}{l}{} &
   &
  \multicolumn{1}{l}{} &
  \multicolumn{1}{l}{} &
   &
  \multicolumn{1}{l}{} &
  \multicolumn{1}{l}{} &
   &
  \multicolumn{1}{l}{} &
  \multicolumn{1}{l}{} \\
 &
  \multirow{3}{*}{3} &
  \multirow{3}{*}{12.2} &
   &
  8 &
  8.5 &
   &
  7 &
  52.8 &
   &
  RW2 &
  -128.9 \\
 &
   &
   &
   &
  12 &
  -65.5 &
   &
  15 &
  -82.5 &
   &
  CRW2 &
  -124.9 \\
 &
   &
   &
   &
  24 &
  57.5 &
   &
  20 &
  55.8 &
   &
  \multicolumn{1}{l}{} &
  \multicolumn{1}{l}{} \\ \bottomrule
\end{tabular}%
}
\caption{Comparison of predictive information criteria between our physics-informed Bayesian state-space models and various Bayesian smoothing techniques using INLA on the simulated data. For B-splines, the knots denotes number of equi-spaced knots for the spline basis. For smoothing using random walks, RW2 model assumes independent second order increments and CRW2 denotes continuous time random walks on second order increments.}
\label{tab:waic}
\end{table}

In each of the above simulated experiments, we report data analysis using the hierarchical model \eqref{eq:uni.model} with mechanistic systems Model~101 and Model~111 as given in \eqref{eq:hew101} and \eqref{eq:hew111}, respectively. For Model~101, we see reasonable posterior learning for $G$ and $Q$ in Figure~\ref{fig:post.learn}(a), whereas for Model 111 the parameters appear to be poorly identifiable. Consequently, we see impaired posterior learning when we assign uninformative priors. Here, the mechanistic parameters $\phi = \{G, Q, Q_R, Q_L, \epsilon_L, \epsilon_{LF}, \epsilon_{RF} \}$ appear as functions $(1-\epsilon_L \epsilon_{LF})G$ and $(Q + \epsilon_{LF} Q_L + \epsilon_{RF} Q_R)$. Therefore, while we see relatively poor learning of individual parameters, learning of the aforementioned functions is reasonable. Hence, learning of the latent process is not compromised. We notice reasonable learning for $(Q + \epsilon_{LF} Q_L + \epsilon_{RF} Q_R)$ as opposed to weaker learning of the individual parameters $Q$, $Q_R$, $Q_L$, $\epsilon_L$, $\epsilon_{LF}$ and $\epsilon_{RF}$. Figure~\ref{fig:post.learn}(b) depicts how more informative priors for the unidentified mechanistic parameters, constructed by considering narrower uniform distributions, positively impact their learning. Specifically, the deviation of the posterior median from their corresponding true values tends to decrease in the presence of stronger prior information. Hence, strongly informative priors are necessary if estimates of these individual parameters are desired. However, deviation of the posterior median estimates of $Q' = Q+ \epsilon_{LF} Q_L + \epsilon_{RF} Q_R$ displays almost no change for the different priors. Investigators studying exposure assessments are interested in estimating these functional forms instead of individual parameters. For example when modeling dynamics of infectious respiratory aerosols, the quantities $Q/V$ in Model~101 or $(Q + \epsilon_{LF} Q_L + \epsilon_{RF} Q_R)/V$ for Model~111 correspond to aerosol removal rates that are important in analyzing air changes per hour (ACH), which, in turn, can inform about probability of infection spread.

We also assess the state-space model's effectiveness in capturing the latent process at the observed time points using the posterior predictive distribution (\ref{eq:smooth1}). Furthermore, for unobserved time points we smooth the latent and observed concentrations using \eqref{eq:smooth1}~and~\eqref{eq:smooth2}, respectively. Subsequently, we compare the performance of each model with different Bayesian semiparametric regressions that do not incorporate information from the underlying mechanistic system. In particular, we considered methods such as B-splines, natural cubic splines and random walk models of order 2 estimated using Integrated Nested Laplace Approximation \citep[INLA:][]{inla1, inla2}. We specifically consider a continuous random walk model on second order increments \citep[denoted CRW2, as described in Section~3.5 of][]{gmrf_book}. Table~\ref{tab:waic} presents overall model comparisons using Watanabe-Akaike Information Criterion \citep[WAIC:][]{waicsingular, waic} as implemented in the \emph{LaplacesDemon} \citep{laplacesdemon} package for the \texttt{R} statistical computing environment \citep{r_env}. We report these scores for the different experimental scenarios and compare our proposed hierarchical model with semiparametric smoothing using B-splines, cubic splines, independent second order increments \citep[denoted RW2, as described in Section~{3.4} of][]{gmrf_book} and CRW2. Table~\ref{tab:waic} shows that while our porposed model significantly outperforms semiparametric smoothing models in out of sample forecasting, the overall model fit as summarized by WAIC between these methods are much more competitive, and in some cases significantly better, than our proposed model. For the single cycle data, WAIC scores for our proposed model with Model 101 are considerably lower than all other methods, while with Model 111, all of the models are competitive in a single cycle. On the other hand, the two random walk models produce significantly lower WAIC scores than the others. 

\begin{figure}
     \centering
     \subfigure[Smoothing on untransformed data]{\label{fig:inla1}\includegraphics[width=0.49\textwidth]{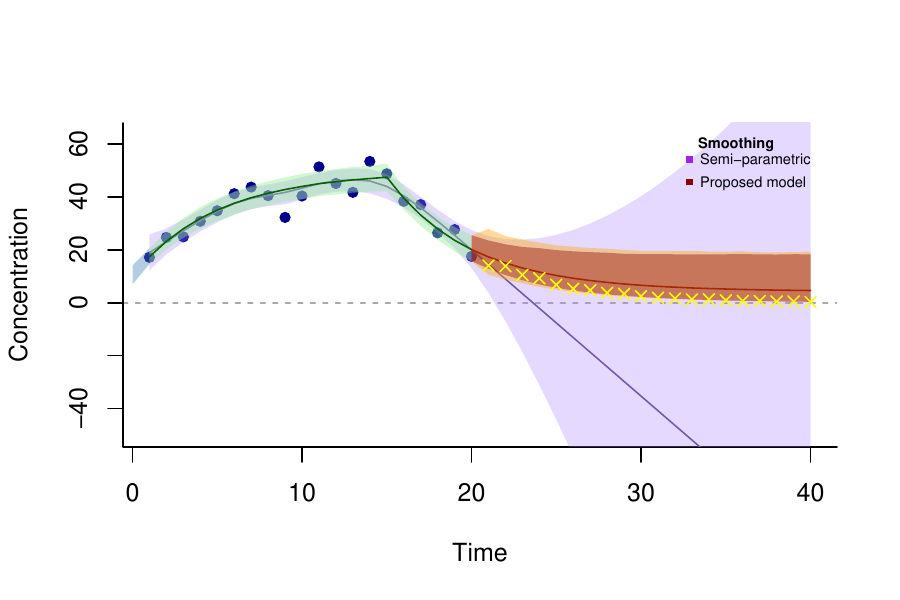}}
     \subfigure[Smoothing on $\log$ transformed data]{\label{fig:inla2}\includegraphics[width=0.49\textwidth]{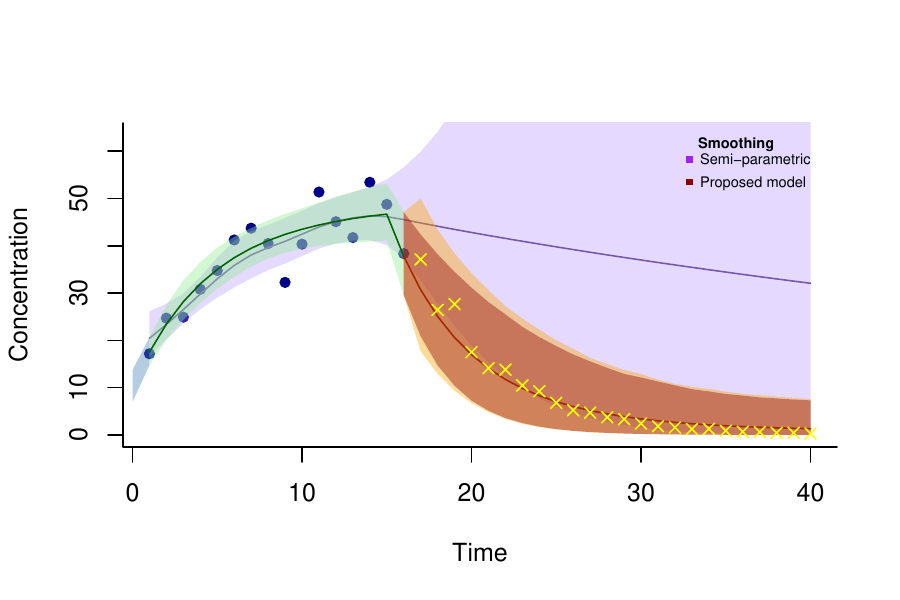}}
     \caption{Prediction and forecasting performances for our hierarchical model \eqref{eq:uni.model} and CRW2-based smoothing on the simulated data for the first $t_N$ time points (marked in blue) with $T_0 = 15$. In Figures~\ref{fig:inla1}~and~\ref{fig:inla2} we used $t_N = 20$ and $t_N = 16$ respectively.}
     \label{fig:bsp.vs.ssm}
\end{figure}

That RW2 and CRW2 excel in terms of WAIC is likely attributable to their interpolation capabilities surrounding the availability of significantly more data in the 3-cycle experiments. In fact, we see a roughly 21\% increase in the residual sum of squares for \eqref{eq:uni.model} over RW2. However, we caution against overstating the excellence of these random walk models that have no mechanistic information. As seen in Figure~\ref{fig:bsp.vs.ssm}, in the absence of mechanistic information, forecasting suffers significantly with these random walk models. Furthermore, the aforementioned reduction in the residual sum of squares should warn investigators against over-fitting. Finally, even if these models estimate concentration levels efficiently, they do not inform about the mechanistic process parameters that govern the underlying physics.

Figure~\ref{fig:bsp.vs.ssm} shows a simple out-of-sample analysis comparing our physics-informed model with semiparametric smoothing. The latter, besides delivering wider uncertainty bands, tends to poorly estimate the trajectory compared to the former. The purple line and the band shows the trajectory of the concentrations fitted using semiparametric smoothing along with the uncertainty around it, whereas the red line and the band show the trajectory and associated uncertainty for the out-of-sample points from our proposed physics-informed state-space model. The yellow crosses indicate the out-of-sample data beyond $t_N = 20$ in Figure~\ref{fig:inla1} and $t_N = 16$ in Figure~\ref{fig:inla2}. In Figure~\ref{fig:inla1}, semiparametric smoothing even forecasts negative concentrations if the data is not appropriately transformed. Figure~\ref{fig:inla2} shows that, even under a suitable transformation, forecasts from semiparametric smoothing are sensitive to the time when the data becomes unavailable.

\section{Analysis of rail car experiment}\label{sec:data_analysis}

\cite{DasRam} have collected substantial concentration data based upon designed experiments with different engineering controls, where each experiment consists of exactly three cycles as described in Section~\ref{sec:experiment}. Here, we do not consider the heterogeneity in ventilation patterns created due to the directional flow of aerosols. Instead, under the ``well-mixed room'' assumption, we focus on modeling the data for one experimental run measured at one location inside the rail car. Due to unavailability of expert prior information on the mechanistic parameters, we only consider the calibration procedures in \cite{hewett1} to construct priors for the relevant parameters. 

\subsection{Noise calibration by mechanistic variance evolution}\label{subsec:noisecalib}
Experimental data from aerosol concentrations contain considerable amounts of noise, which often exceeds the capabilities of a statistical model equipped with uniform error variance over time to quantify uncertainty. As we expect the aerosol concentration measurements to appear in varying scales across the duration of an experimental cycle, we consider the influence of mechanistic factors on the evolution of the error variance over time. A simple yet effective approach to address this is to introduce a dynamic $v_t = v_t(\phi)$ scale factor in the variance of $\omega_t$ in transition equation in \eqref{eq:hier}. This scale factor depends on $\phi$ since the mechanistic parameters dictate how the data are generated and, hence, how its variability evolves. Therefore, we can modify the transition equation in \eqref{eq:uni.model} as
\begin{align}\label{eq:dynvar}
\begin{split}
    C_t &= \left(1 - \frac{\Delta}{V} Q\right) C_{t-1} + \frac{\Delta}{V} G_t + v_t \omega_t \\
    v_t &= H_t \, v_{t-1} \\
    H_t &= (1 + \alpha) 1_\mathcal{G}(t) + \beta (1 - 1_\mathcal{G}(t)) \\
    \{\alpha, \beta\} & \sim p(\alpha) \, p(\beta)\;,
\end{split}
\end{align}
where $1_\mathcal{G}(t) = 1$ when $G_t = G$ (i.e., the particle generator is in place) and $1_\mathcal{G}(t) = 0$ when $G_t = 0$ (no generation). With $\alpha > 0$, we model the error variances in the transition equation to change in a multiplicative fashion - increasing as long as the particle generator is on, and decreasing with $0 < \beta < 1$ after the generator is turned off. These modifications are applicable to \eqref{eq:obs.transn}. Since the transition equation in \eqref{eq:obs.transn} is derived from a finite difference approximation of the original system, we may not be able to easily characterize the error distribution from a transformation of $\{C_t\}_{t \geq 1}$ in the transition equation while also maintaining an appropriate first-order Markov dependence.  The above model is implemented in the computing environment described in Section~\ref{subsec:computation}. Posterior inference reported here is based on 5,000 MCMC samples after discarding the first 5,000 burn-in samples.

\begin{figure}[t]
     \centering
     \subfigure[Time-independent process evolution errors]{\label{fig:badfit}\includegraphics[width=0.7\textwidth]{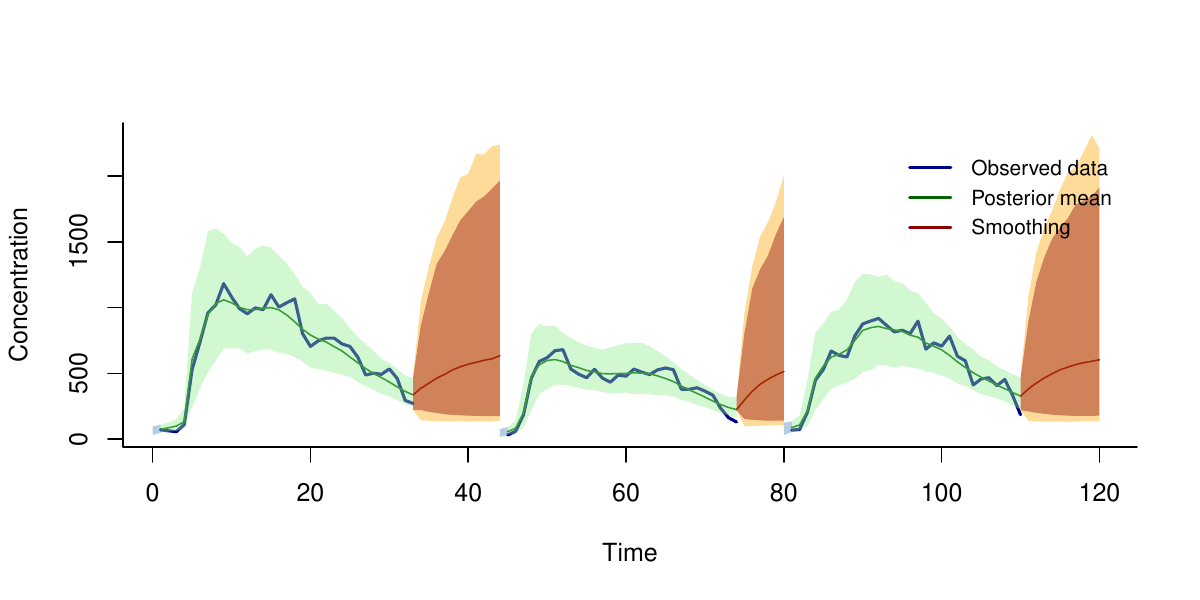}}
     \subfigure[Time-dependent process evolution errors]{\label{fig:dynvarfit}\includegraphics[width=0.7\textwidth]{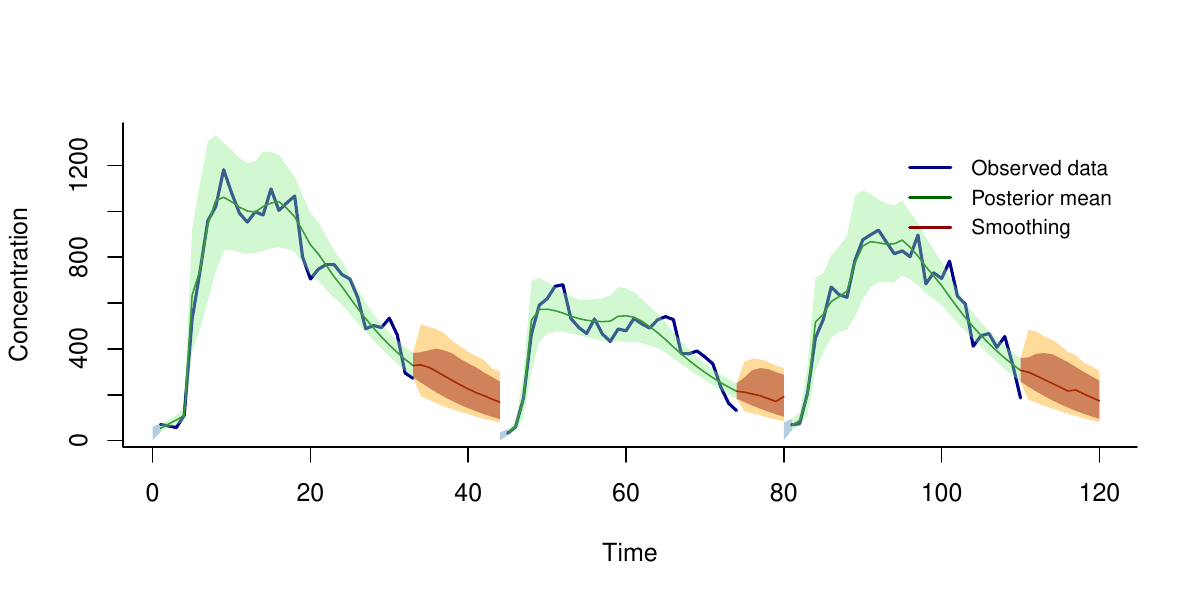}}
     \caption{Smoothing and forecasting on noisy experimental data by models with and without evolution of error variances over time.}
     \label{fig:dynvar}
\end{figure}

In the current context, we find that our modified model as in \eqref{eq:dynvar} for the latent process evolution adequately provides robust analysis and a similar modification in the observation equation is unnecessary. This choice is corroborated by restrictions on the support of the error variance imposed by $\{C_t\}_{t \geq 1}$. As concentrations are positive quantities, we modeled the transition errors using a lognormal distribution. As a result, when the noisy experimental data is fitted with a model with time-independent transition errors, the parameter estimates in the lognormal distribution yield inaccurate and unreasonably wide uncertainty bands for smoothing and forecasting. Considering time-dependent errors in the transition equation resolves this problem by suitably calibrating the errors informed by the aerosol generation status provided by the mechanistic system. This enriches Bayesian melding of mechanistic information and the statistical model.  Figure~\ref{fig:dynvar} presents these comparisons.

\subsection{Results}
In addition to obtaining uncertainty quantification of the underlying process for the particle concentration using the dynamic variance approach, as elaborated in Section~\ref{subsec:noisecalib}, our Bayesian model also delivers samples from the posterior distribution of the mechanistic parameters. Table~\ref{tab:post_field} presents posterior summaries of the parameters appearing in Model~111 that includes particle generation rate, different ventilation rates and filtration efficiencies. 
\begin{table}[t]
\centering
\begin{tabular}{crrrc}
\toprule
Parameter & Median & 2.5\% & 97.5\% & Unit \\ \midrule
$G$ & 6863.50 & 2854.02 & 9818.22 & mg/min \\
  $Q$ & 11.41 & 10.05 & 14.36 & m$^3$/min\\
  $Q_L$ & 2.20    & 0.16    & 6.66 &  m$^3$/min \\
  $Q_R$ & 1.67    & 0.17    & 6.34 &  m$^3$/min \\
  $\epsilon_L$ & 0.83 & 0.71 & 0.99 & -\\
  $\epsilon_{LF}$ & 0.66    & 0.51    & 0.95  & - \\
  $\epsilon_{RF}$ & 0.67    & 0.51    & 0.89  &  - \\ \bottomrule
\end{tabular}
\caption{Posterior summary of different mechanistic parameters obtained from the analysis of data from a single railcar experiment. The filtration efficiencies $\epsilon_L$, $\epsilon_{LF}$ and $\epsilon_{RF}$ are unitless quantities and lie between 0 and 1.}
\label{tab:post_field}
\end{table}
The posterior samples of these parameters are crucial for finding the posterior distribution of various quantities that are relevant to a practicing industrial hygienist. For example, we obtain the posterior distribution of the quantity $(Q + \epsilon_{LF} Q_L + \epsilon_{RF} Q_R)/V$ which corresponds to the total removal rate of the particles (in units of min$^{-1}$) by evaluating the quantity at each posterior sample of the parameters involved in the quantity. Based on data analysis from a single experiment, we find the total removal rate of the particles to be approximately 5.68 hr$^{-1}$ with 95\% credible interval (4.64 hr$^{-1}$, 6.90 hr$^{-1}$). Another quantity of interest, the average concentration of particles ($C_{\text{avg}}$) in the rail car is estimated by
\begin{equation*}
    C_{\text{avg}} = \frac{G'}{Q'} \left[ 1 - \frac{1}{(Q'/V)T} \left\{ 1 - \exp \left(- \frac{Q'}{V} T \right) \right\} \right] \;,
\end{equation*}
where $G' = (1-\epsilon_L \epsilon_{LF})G$ and $Q' = Q + \epsilon_{LF} Q_L + \epsilon_{RF} Q_R$. Here, $T$ corresponds to the total duration of the experiment. We find the posterior median of $C_{\text{avg}}$ to be 200.53 with 95\% credible interval (73.90, 286.32). The quantity $C_{\text{avg}}$ is useful for modeling SARS‐CoV‐2 airborne quanta transmission and exposure risk that estimates probability of infection \citep{yan2022risk, DasRam}.

\begin{figure}
    \centering
    \includegraphics[width = 0.6\textwidth]{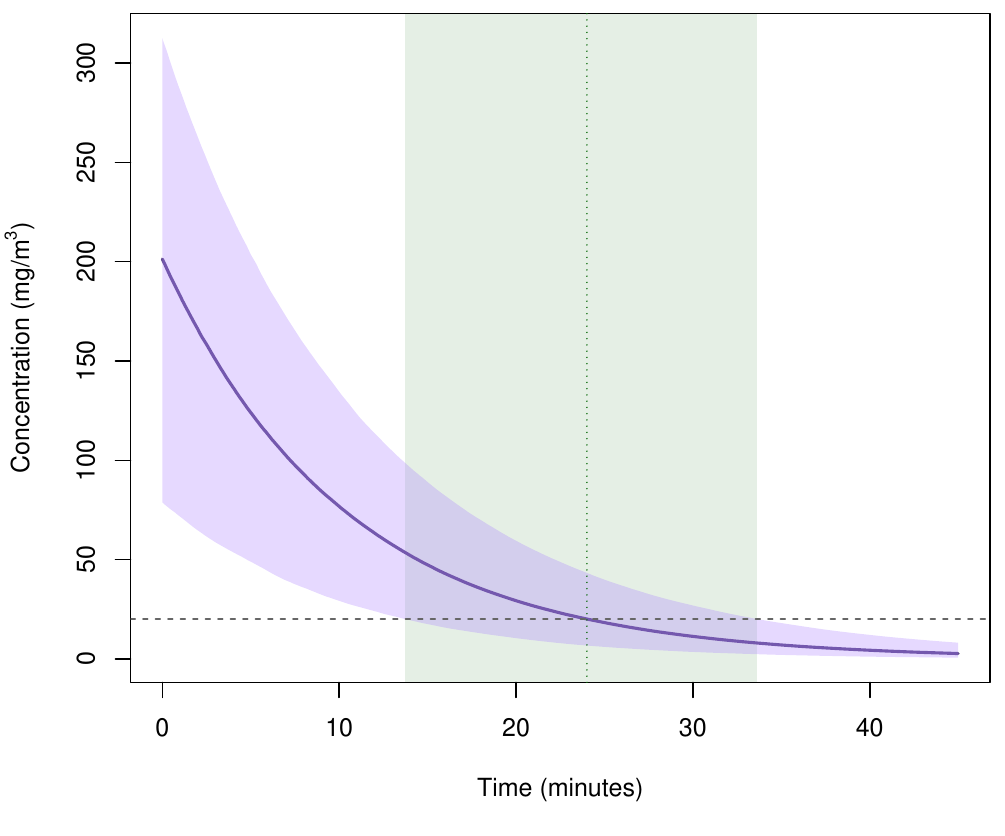}
    \caption{Estimated concentration decay curve (purple) along with its 95\% uncertainty bands. The vertical dotted line denotes the posterior median of time required for the concentration to drop below a certain threshold (horizontal dashed) while and the shaded window denotes its 95\% credible interval.}
    \label{fig:decay}
\end{figure}

Assuming continuous particle generation at a constant rate $G$, the steady-state concentration, defined as the limit of the concentration $C(t)$ as $t \to \infty$, is given by $C_{\infty} = G'/Q'$ for Model 111. From our analysis, we find the steady-state concentration to be approximately 245.95 mg/m$^3$ with a a 95\% credible interval of (95.29 mg/m$^3$, 324.71 mg/m$^3$). Moreover, in the absence of particle generation, we can estimate the time taken by the ventilation system to reduce the particle concentration from $C_1$ mg/m$^3$ to $C_2$ mg/m$^3$ by $(V/Q') \log(C_1/C_2)$ minutes. Industrial hygienists may use this quantity to understand the capacity of the ventilation system by studying the time required for the particle concentration to drop under a specified threshold indicating low contamination. Figure~\ref{fig:decay} illustrates the posterior concentration decay curve with $C_{\infty}$ as the initial particle concentration along with the estimated time to reach a threshold indicating a low contamination level. For example, if the threshold is specified as 10\% of the estimated steady-state concentration, then the ventilation system requires around 24 minutes to bring the concentration from its steady-state down to the threshold with a 95\% credible interval of (15 mins, 34 mins).

\section{Discussion}\label{sec:discussion}

We show that while certain mechanistic parameters may not be well identified by the field data, the latent concentration process is effectively estimated. Assimilating mechanistic systems in our data analysis framework yields especially pronounced benefits in forecasting performance over flexible semiparametric smoothing techniques that do not assimilate such systems, while all these methods may indicate adequate goodness of fit. 

Practicing industrial hygienists and occupational exposure modelers are often encumbered in their conclusions by the challenges in designing controlled experiments that can generate data consistent with mechanistic processes posited to be generating concentrations in ideal physical settings. When faced with the task of estimating physical parameters and concentrations from experimental data, exposure modeler's generate synthetic concentrations from the mechanistic model for different input parameters and essentially ``eye-ball'' which values of the input yields concentration values closest to those from the experimental data. Even if formal metrics based upon ``distances'' between mechanistically generated data and field observations, as is done nowadays in some machine learning methods, are used to choose ``optimal'' values of the mechanistic parameters, such methods preclude uncertainty quantification in mechanistic inputs and noise in measurements. The tools we have described here offers a statistically justifiable way to estimate the mechanistic parameters. We achieve two practical goals of occupational exposure analysis: (i) estimating mechanistic parameters combining information from prior mechanistic considerations and from experimental data; and (ii) meld or assimilate mechanistic models with statistical models to provide improved predictions for concentrations, including better forecasts into the future, while also offering full uncertainty quantification.

Extensions of our models are possible in different directions. For example, in controlled experiments it is typical of photo detector particle counters to offer particle counts in several size ranges that are expected to be correlated. This stokes the possibility of jointly modeling the particle sizes in the process. For $p$ different size-ranges, let $Y_t$ and $C_t$ be $p$-variate observed and latent concentrations at time $t$. A multivariate framework for the one-box model is
\begin{align}\label{eq:multi.model}
\begin{split}
    g(Y_t) \mid C_t, \beta, \sigma^2_\upsilon &\sim \mathcal{N}_p(g(C_t) + X_t^\top \beta, \Sigma) \\
    C_t \mid \phi, m_\omega, \sigma^2_\omega & \sim \mathrm{ShiftedLMN} (A_t(\phi, \Delta_t)^\top \, C_{t-1} + B_t(\phi, \Delta_t); m_{\omega}, \sigma^2_\omega I_p) \\
    \{\phi, \beta, \sigma^2_\upsilon, m_{\omega}, \sigma^2_\omega \} &\sim p(\phi)\, p(\beta \given \sigma^2_\upsilon)\, p(\sigma^2_\upsilon)\, p(m_\omega)\, p(\sigma^2_\omega)\;,
\end{split}
\end{align}
where $A_t(\phi, \Delta_t) = 1_p - (Q + \epsilon_{LF} Q_L + \epsilon_{RF} Q_R)\Delta_t/V$ and $B_t(\phi, \Delta_t) = (1-\epsilon_L \epsilon_{LF})G 1_\mathcal{G}(t)\Delta_t/V$ are calculated using element wise operations applied to the $p$-variate parameters in $\phi$. Here, $X+\theta$ is distributed as $\mathrm{ShiftedLMN}(\theta; \mu, V)$ if $\log X$ is distributed as multivariate normal with mean $\mu$ and covariance matrix $V$ for some $\theta \in \mathbb{R}^p$. Further investigations into the dependence structure among size-specific particle concentrations is open to future investigations as are questions on the structure of $\Sigma$ in \eqref{eq:multi.model} and its effects of inference. 

While the current analysis advocates delving in the mechanistic equations as a part of the model building exercise, we recognize that such luxuries may be precluded by more complex models in other application. In this regard, stochastic emulators such as Gaussian processes \citep[]{monteiro2014tch} are widely employed to conduct such inference. We have not undertaken a comprehensive comparison with such methods in this paper and recognize them as viable options in our current setting. Such approaches will enable broader scope of mechanistic explorations in occupational exposure field data settings and comprise an area of future research.  

\section*{Conflict of Interest}
None declared.

\section*{Funding}
Banerjee, Ramachandran and Pan were supported, in part, from the National Institute of Environmental Health Sciences (NIEHS) R01ES030210. Banerjee also acknowledges support from NIEHS R01ES027027, the National Institute of General Medical Science (NIGMS) R01GM148761, and the Division of Mathematical Sciences (DMS) of the National Science Foundation 
2113778.


\bibliographystyle{plainnat}
\bibliography{refs}

\end{document}